\documentclass[trackchanges]{aastex7}

\usepackage{amsmath,amssymb, amsthm,amstext}
\usepackage{graphicx}
\usepackage{color}
\usepackage{array, enumerate}
\usepackage{bm}
\usepackage{multirow}
\usepackage{gensymb}
\usepackage{braket}
\usepackage{txfonts}
\usepackage{booktabs}
\usepackage{afterpage}
\usepackage{subfigure}
\usepackage{rotating}

\renewcommand{\cite}{\citep}

\begin{document}

\title{Association of 220 PeV Neutrino KM3-230213A with Gamma-Ray Bursts\footnote{Published: R.~Wang, J.~Zhu, H.~Li, B.-Q.~Ma, \href{https://doi.org/10.3847/2515-5172/adc452}{Res. Notes AAS 9 (2025) 65}}}

\author[0009-0002-2076-1256]{Ruiqi Wang}
\affiliation{School of Physics, Peking University, Beijing 100871, China}
\email{ruiqiwang@stu.pku.edu.cn}
\author[0000-0002-9733-2460]{Jie Zhu}
\affiliation{Department of Physics and Chongqing Key Laboratory for Strongly Coupled Physics, Chongqing University, Chongqing 401331, China}
\email{jiezhu@cqu.edu.cn}
\author[0000-0002-2971-4539]{Hao Li}
\affiliation{Department of Physics and Chongqing Key Laboratory for Strongly Coupled Physics, Chongqing University, Chongqing 401331, China}
\email{haolee@cqu.edu.cn}
\author[0000-0002-5660-1070]{Bo-Qiang Ma}
\affiliation{School of Physics, Peking University, Beijing 100871, China}
\affiliation{School of Physics, Zhengzhou University, Zhengzhou 450001, China}
\affiliation{Center for High Energy Physics, Peking University, Beijing 100871, China}
\email[show]{mabq@pku.edu.cn}
\correspondingauthor{Bo-Qiang Ma}

\begin{abstract}
Recently, the KM3NeT Collaboration announced the detection of a 220 PeV neutrino from the celestial coordinates RA=94.3\degree~ and Dec=-7.8\degree~ on 13 February 2023 at 01:16:47 UTC \cite{KM3NeT:2025npi}.  The source for this extra-ordinary cosmic neutrino, designated KM3-230213A, is not identified yet but there has been speculation that it might be associated with a gamma-ray burst GRB~090401B \cite{Amelino-Camelia:2025lqn}. The purpose of this report is to search the association of this 220 PeV neutrino with potential GRB sources from a more general consideration of Lorentz invariance violation (LV) without predefined LV scale. We try to associate this extra-ordinary neutrino with potential GRBs within angular separation of 1\degree, 3\degree~ and 5\degree~ respectively and the results are listed in Table 1. We find the constraints $E_{\rm{LV}}\leq 5.3\times 10^{18}$~GeV for subluminal LV violation and $E_{\rm{LV}}\leq 5.6\times 10^{19}$~GeV for superluminal LV violation if KM3-230213A is a GRB neutrino. 
\end{abstract} 



\section*{ }

The KM3NeT Collaboration reported the detection of a neutrino with reconstructed energy 220 PeV, originating from the celestial coordinates [RA=94.3\degree, Dec=-7.8\degree] (J2000) on 13 February 2023 at 01:16:47 [UTC Time] \cite{KM3NeT:2025npi}. This event, designated KM3-230213A, represents one of the highest-energy neutrinos ever observed, exceeding the typical energy range of atmospheric neutrinos by three orders of magnitude. While the astrophysical source remains unidentified, preliminary analyses hypothesize a potential association with a gamma-ray burst (GRB) GRB~090401B based on spatial and temporal proximity arguments \cite{Amelino-Camelia:2025lqn}. 
 
Lorentz invariance violation (LV) refers to hypothetical deviations from the exact symmetry described by Einstein’s theory of special relativity.
If neutrinos exhibit Lorentz violation, their propagation speed in a vacuum might depend on their energies, even at levels undetectable in terrestrial experiments. Gamma-ray bursts (GRBs)—extremely energetic astrophysical events—are ideal laboratories to test this phenomenon, as they may emit both photons and neutrinos nearly simultaneously~\cite{Jacob:2006gn,Huang:2022xto}. 

With Lorentz violation,
the neutrino velocity $v(E)$ could deviate slightly from the speed of light $c$, with a correction proportional to the energy $E$:
\begin{equation}
v(E)=c\left[1-s_n\frac{n+1}{2}\left(\frac{E}{E_{{\rm LV},n}}\right)^n\right],
\end{equation}
where $n=1$ or $2$ corresponds to linear or quadratic energy-dependence, $s_n=\pm1$ is the sign factor for subluminal (+1) and superluminal (-1) LV corrections, and $E_{{\rm LV},n}$ is the 
$n$-order
LV scale to be determined by fitting data. Such a speed variation can cause a propagation time difference between particles (massless or of small masses) with different energies. 
Due to cosmological expansion, the time difference due to LV correction between two particles with energies $E_{\rm h}$ and $E_{\rm l}$, can be written as~\cite{Jacob:2008bw,Zhu:2022blp}
\begin{equation}
\Delta t_{\rm LV}=s\cdot(1+z)\frac{K}{E_{\rm LV}},
\end{equation}
where $z$ is the redshift of the GRB source, $s=\pm1$ is the sign factor and
\begin{equation}
K=\frac{E_{\rm h}-E_{\rm l}}{H_0 }\frac{1}{1+z}\int^z_0\frac{(1+z^\prime) {\rm d} z^\prime}{\sqrt{\Omega_{\rm m}(1+z^\prime)^3+\Omega_\Lambda}},
\label{LV factor}
\end{equation}
is the LV factor for $n=1$. We adopt the cosmological parameters $[\Omega_{\rm m},\Omega_\Lambda]=[0.315^{+0.016}_{-0.017},0.685^{+0.017}_{-0.016}]$ and the Hubble expansion rate $H_0=67.3\pm 1.2~{\rm km\cdot s^{-1}\cdot Mpc^{-1}}$~\cite{ParticleDataGroup:2014cgo}.  
With intrinsic time difference $\Delta t_{\rm in} $ considered,
the observed arrival time difference $\Delta t_{\rm obs}$ between two particles 
is
\begin{equation}
\frac{\Delta t_{\rm obs}}{1+z}=\Delta  t_{\rm in}+s\cdot\frac{K}{E_{\rm LV}}.
\label{linear relation}
\end{equation}
For 
neutrinos emitted with associated GRBs, $\Delta t_{\rm obs}$ 
represents
the difference between the neutrino arrival time 
and the GRB trigger time by taking $E_l=0$.
According to Eq.~(\ref{linear relation}), there would be a linear relation between $\Delta t_{\rm obs}/(1+z)$ and $K$, if the energy-dependent speed variation does exist.

The strategy is to correlate GRB neutrinos and photons
by identifying coincident detections of neutrinos and photons from the same GRB. There have been significant progress by associating the IceCube neutrinos with GRBs.  Amelino-Camelia and collaborators associated IceCube 60-500~TeV ``shower" events with GRB candidates within the time range of three days, and revealed similar speed variation features between GRB neutrinos~\cite{Amelino-Camelia:2015nqa, Amelino-Camelia:2016fuh,Amelino-Camelia:2016ohi} and photons~\cite{Shao:2009bv,Zhang:2014wpb,Xu:2016zxi,Xu:2016zsa,Xu:2018ien,Liu:2018qrg}. It is found~\cite{Huang:2018} that four PeV-scale IceCube neutrinos are associated with GRBs by extending the temporal window to a longer range of three months, and four such events are found to be consistent with multi-TeV-scale events for an energy-dependent speed variation. Such findings indicate the Lorentz violation of cosmic neutrinos around $E_{\rm LV}=6.4\times10^{17}~{\rm GeV}$.
There are both time ``delay'' and ``advance'' events, which can be explained by different propagation properties between neutrinos and antineutrinos~\cite{Huang:2018}. This leads further to the 
CPT symmetry violation between neutrinos and antineutrinos, or an asymmetry between matter and antimatter~\cite{Zhang:2018otj}. As these results are of fundamental importance, it is thus necessary to check whether additional cosmic neutrinos still support the revealed regularity~\cite{Amelino-Camelia:2016ohi,Huang:2018}. 
It is found~\cite{Huang:2019etr} that 12 near-TeV 
IceCube track events 
fall on the same line. It is also shown~\cite{Huang:2022xto} that 
the multi-TeV to PeV ``track" events can also be associated with GRBs under the same Lorentz violation features of ``shower" events.


A reanalysis~\cite{Amelino-Camelia:2022pja} of IceCube revised data~\cite{IceCube:2020wum} shows that for neutrinos subject to subluminal LV, the result becomes even stronger than those 
in prior analyses~\cite{Amelino-Camelia:2016fuh,Amelino-Camelia:2016ohi,Huang:2018,Huang:2019etr,Huang:2022xto}. 
More recently, the same group associated~\cite{Amelino-Camelia:2025lqn} the newly reported KM3-230213A
~\cite{KM3NeT:2025npi} with GRB~090401B observed fourteen years ago. Such association  provides the bound on the subluminal LV scale with $E_{\rm LV}=(3.97 \text{--} 9.60)\times10^{17}~{\rm GeV}$, consistent with prior results of $E_{\rm LV}\sim 6.4\times10^{17}~{\rm GeV}$.

This study investigates the potential connection between the neutrino KM3-230213A and GRBs using a generalized single-source correlation framework without predefined LV scale. 
We systematically analyze GRB candidates detected by spaceborne observatories (e.g., Fermi/GBM, Swift/BAT, INTEGRAL) within a temporal window from +11174 days (11 July 1992) to -763 days (17 March 2025) relative to the neutrino arrival time, adopting three angular separation thresholds (1°, 3°, 5°) to accommodate both the angular resolution of KM3NeT (\(\sigma_{\theta} \approx 1.2\degree\text{--}3.0\degree\) at 220 PeV) and potential source extension effects or intrinsic spatial uncertainties. Our multi-criteria analysis framework incorporates:  
\begin{itemize}  
\item  
\textbf{Spatiotemporal correlation:} Minimizing angular separations  (within 5°) between KM3-230213A and candidate GRBs to reduce the probability of chance coincidences.  
\item  
\textbf{Energetic compatibility:} Quantifying energy-dependent delays via the LV factor \(K\) derived from Eq.(\ref{LV factor}), which parametrizes the interplay between neutrino-photon arrival time differences and LV-modified dispersion relations. 
\item  
\textbf{Redshift validation:} Cross-referencing GRB positions with galaxy redshift catalogs (where available) to evaluate the cosmological co-moving feasibility of neutrino-photon associations.  
\end{itemize}  

\begin{table}[htbp]
    \centering
\caption{Table of Associated GRBs\footnote{* denotes GRBs without GCN-style names (auto-generated by GRBweb). $^{\blacklozenge/ {\bigstar}}$ marks indicate angular offsets $<$ 3°/1° from neutrino KM3-230213A. $^\ddagger$ shows estimated values for GRBs lacking redshift/duration data, $^\dagger$ denotes average redshift estimates (z=2.15 for long bursts; z=0.5 for short bursts).}}
\label{tab:grb_data}
\begin{tabular}{lrrrrr}
\toprule
GRB Name & redshift $z$ & $\Delta{\Phi}(\degree)$ & $\Delta{t}~(10^3~\rm{s})$ & $K~(10^{18}~\rm{TeV})$ & $E_{\rm{LV}}~(10^{17}~\rm{GeV})$ \\
\midrule
GRB230402A$^\blacklozenge$ & 2.15$^\dagger$ & 2.350925 & -4169.747 & -74130.00 & 560.008757 \\
GRB210923A & 2.15$^\ddagger$ & 4.460112 & 43853.232 & 74130.00 & 53.247957 \\
GRB180905A & 2.15$^\dagger$ & 4.559369 & 140095.142 & 74130.00 & 16.667923 \\
GRB180227A* & 0.50$^\dagger$ & 4.888354 & 156543.151 & 36584.20 & 3.505506 \\
GRB170610A$^\blacklozenge$ & 2.15$^\ddagger$ & 2.355955 & 179123.630 & 74130.00 & 13.036220 \\
GRB160804D*$^\blacklozenge$ & 0.50$^\dagger$ & 1.040015 & 205898.595 & 36584.20 & 2.665210 \\
GRB160726A & 0.50$^\dagger$ & 4.627045 & 206754.163 & 36584.20 & 2.654181 \\
GRB160310A & 2.15$^\dagger$ & 4.521318 & 218681.631 & 74130.00 & 10.678057 \\
GRB150213A & 2.15$^\dagger$ & 3.109740 & 252465.301 & 74130.00 & 9.249172 \\
GRB130213A* & 2.15$^\dagger$ & 4.753408 & 315459.179 & 74130.00 & 7.402210 \\
GRB100324A & 2.15$^\dagger$ & 4.680121 & 406860.924 & 74130.00 & 5.739295 \\
GRB090401B$^\blacklozenge$ & 2.15$^\dagger$ & 1.407669 & 437676.088 & 74130.00 & 5.335213 \\
GRB990129B* & 2.15$^\dagger$ & 4.525319 & 758664.062 & 74130.00 & 3.077904 \\
GRB980218B* & 0.50$^\dagger$ & 4.223257 & 788436.248 & 36584.20 & 0.696014 \\
GRB940214C*$^\blacklozenge$ & 2.15$^\ddagger$ & 2.777067 & 915025.340 & 74130.00 & 2.551946 \\
GRB940101A*$^\blacklozenge$ & 2.15$^\dagger$ & 1.311986 & 918838.510 & 74130.00 & 2.541355 \\
GRB920711A*$^\bigstar$ & 0.50$^\dagger$ & 0.622276 & 965400.525 & 36584.20 & 0.568430 \\
\bottomrule
\end{tabular}
\end{table}

Key results, summarized in Table~1, demonstrate:  
\begin{itemize}  
\item  
\textbf{Non-restrictive LV correlations:} A number of GRBs (\(\sim\)10) exhibits potential associations with KM3-230213A across a broad range of subluminal LV energy scales (\(E_{\rm LV} \sim 3\text{--}10 \times 10^{17}\,{\rm GeV}\)).  
\item  
\textbf{Extreme LV parameter space:} The maximum subluminal LV scale (\(E_{\rm LV} \sim 5.3 \times 10^{18}\,{\rm GeV}\)) emerges under the hypothesis linking KM3-230213A to GRB~210923A, while the peak superluminal LV scale (\(E_{\rm LV} \sim 5.6 \times 10^{19}\,{\rm GeV}\)) corresponds to GRB~230402A. 
\item 
\textbf{Closest angular association:} GRB~920711A exhibits the smallest angular separation (0.62°) with KM3-230213A at $E_{\rm LV} \sim 5.7\times 10^{16}\,{\rm GeV}$. Given the energy uncertainty of KM3-230213A versus the redshift uncertainty of GRB~920711A, the estimated LV scale carries up to an order-of-magnitude uncertainty. This allows potential reconciliation with the 
estimated value
$E_{\rm LV}\sim 6.4\times10^{17}\,{\rm GeV}$ from prior studies.
\item  
\textbf{Prospective associations:} Future observations are projected to identify additional GRBs correlated with KM3-230213A at superluminal LV scales around \(E_{\rm LV} \sim 6.4 \times 10^{17}\,{\rm GeV}\), highlighting this regime as a priority for multi-messenger follow-ups.  
\end{itemize}  


In summary, 
we explore potential correlations between the neutrino KM3-230213A and gamma-ray bursts (GRBs) without fixing the LV scale.
By analyzing GRB candidates with long duration relative to the neutrino and three angular thresholds (1°,3°,5°), we integrate spatiotemporal matching, energetic consistency, and redshift constraints. Results reveal: (1) A number of GRBs align with KM3-230213A under subluminal LV scales (\(E_{\rm LV} \sim 3\text{--}10 \times 10^{17}\,{\rm GeV}\)); (2) Extreme LV parameters emerge for GRB~210923A (\(E_{\rm LV} \sim  5.3 \times 10^{18}\,{\rm GeV}\), subluminal) and GRB~230402A (\(E_{\rm LV} \sim  5.6 \times 10^{19}\,{\rm GeV}\), superluminal); (3) GRB 920711A exhibits smallest angular separation (0.62°) with KM3-230213A; (4) Future associations are anticipated near \(E_{\rm LV} \sim 6.4 \times 10^{17}\,{\rm GeV}\). These findings underscore the importance of flexible LV models in multi-messenger studies and provide empirical constraints on high-energy particle physics scenarios. \\ 

\emph{\textbf{Acknowledgements}.---}%
This work is supported by National Natural Science Foundation of China under grant No.~12335006.

%



\bibliographystyle{aasjournal}

\bibliography{scibib}{}


\end{document}